\documentclass{article}
\usepackage{ijcai20}
\usepackage{multirow}
\usepackage[caption=false,font=footnotesize]{subfig}
\usepackage{times}
\usepackage{latexsym}
\usepackage{soul}
\usepackage{url}
\usepackage[hidelinks]{hyperref}
\usepackage[utf8]{inputenc}
\usepackage[small]{caption}
\usepackage{graphicx}
\usepackage{amsmath,amssymb,amsfonts}
\usepackage{amsthm}
\usepackage{booktabs}
\usepackage{algorithm}
\usepackage{algorithmic}
\usepackage{xcolor}

\newtheorem{problem}{Problem}

\newcommand{\M}{HDGNN}

\title{A Heterogeneous Dynamical Graph Neural Networks Approach to Quantify Scientific Impact}

\author{
Fan Zhou$^1$
\and
Xovee Xu$^1$
\and
Ce Li$^1$
\and
Goce Trajcevski$^2$
\and
Ting Zhong$^1$
\And
Kunpeng Zhang$^3$
\affiliations
$^1$University of Electronic Science and Technology of China, Chengdu, China\\
$^2$Iowa State University, IA, USA\\
$^3$University of Maryland, College park, MD, USA
\emails
fan.zhou@uestc.edu.cn,
xovee@live.com, 
ce.lc@outlook.com,
gocet25@iastate.edu,  
zhongting@uestc.edu.cn, 
kpzhang@umd.edu
}

\begin{document}

\maketitle

\begin{abstract}


Quantifying and predicting the long-term impact of scientific writings or individual scholars has important implications for many policy decisions, such as funding proposal evaluation and identifying emerging research fields. In this work, we propose an approach based on  Heterogeneous Dynamical Graph Neural Network (HDGNN) to explicitly model and predict the cumulative impact of papers and authors. HDGNN extends heterogeneous GNNs by incorporating temporally evolving characteristics and capturing both structural properties of attributed graph and the growing sequence of citation behavior. HDGNN is significantly different from previous models in its capability of modeling the node impact in a  dynamic manner while taking into account the complex relations among nodes. Experiments conducted on a real citation dataset demonstrate its superior performance of predicting the impact of both papers and authors. 


\end{abstract}

\section{Introduction}
\label{sec:introduction}

The pace of growth 
of the body 
of scientific research has been rapidly increasing in recent years. For example, the number of records in DBLP\footnote{\url{https://dblp.uni-trier.de/statistics/recordsindblp}} has increased from 2,486,800 in 2013, to 4,893,893 in 2019; according to the AI index report 2019\footnote{\url{https://hai.stanford.edu/ai-index/2019}}, the number of peer-reviewed AI publications has increased by 300\% between 1998-2018. 
Quantifying the impact of the publications, as well as individual 
scholars/authors is 
an important task in many domains of societal and scientific relevance. For example, funding agencies and research institutes need to deeply understand the current research development -- e.g., discovering frontier ideas, identifying breakthrough topics and productive scholars, seeking well-fitted scientists for defined projects, hiring high-quality faculties~\cite{Fortunato2018} -- for 
improved policy/decision making. The availability of various scientific databases, such as Web of Science, Google Scholar, DBLP and U.S. Patent, provides an unprecedented opportunity to explore the career of scientists and the dynamic evolving process of paper dissemination. However, scientific impacts of scholars and papers can be affected by a variety of factors. For example, a productive researcher may publish a number of papers every year, but the impact of her/his publications may vary 
significantly~\cite{sinatra2016quantifying}. Also, some scientific findings may receive a burst of attention immediately, while 
others may take decades since 
their original publication date~\cite{Ke2015}. 


\noindent \textbf{Existing works}: Quantifying and foreseeing the (impact of) scientific diffusion have been scrutinized by generations of researchers since~\cite{Price1965}. Earlier efforts~\cite{yan2011citation,Acuna2012,wang2013quantifying} primarily focused on extracting indicative features and discovering latent mechanisms that drive the accumulation of citations. 
Features of scholars such as the number of publications, the $h$-index, years since the first publication were used to forecast the future $h$-index in~\cite{Acuna2012}. 
Factors such as 
topical authority and publication venue 
that may increase citations were used to predict the scientific impact in~\cite{dong2016can}. 
Despite certain merits, 
the previous works are limited on the impact predictability due to the confluence of different and sometimes controversial factors~\cite{Clauset2017,Ke2015}, and the difficulty of generalizing the knowledge from one discipline to another. Meanwhile, some implicit but important factors are not fully leveraged in a scientific way, such as academic authority that amplifies author/paper exposure and facilitates grants funding. 
Another line of work predicts the propagation of scientific impact from the perspective of stochastic information dynamics, 
relying on various pattern-recognition based models (e.g., Hawkes process and Poisson process)~\cite{shen2014modeling,cao2017deephawkes}. 
These methods are theoretically solid and demonstrated their advancement, 
particularly for interpretability, but they require longer sequences of observations and 
are unable to fully leverage the complex interactions among authors and papers for 
impacts prediction.

Recent applications of deep neural networks on graph-based data have inspired numerous models for capturing temporal and sequential process of information diffusion, including scientific impact. DeepCas~\cite{li2017deepcas} is a graph-embedding based popularity prediction model, learning the representation of cascade graphs with DeepWalk~\cite{perozzi2014deepwalk} and the diffusion process via recurrent neural networks~\cite{chung2014empirical}. CasCN~\cite{chen2019cascn} exploits the structure of each information cascade by a dynamic graph convolutional network (GCN)~\cite{Kipf2017}, and predicts the size of cascades while taking the directionality of cascades and time decay effects into consideration. 
As a multi-task learning framework, it 
simultaneously predicts the information popularity at the macro-level and the user participation in re-posting at the micro-level
However, these approaches deal with representation learning of homogeneous graphs, which limits their capability of exploiting the information associated with node attributes and complex relations among heterogeneous nodes. Thus, 
incorporating 
meaningful relations among nodes into the information diffusion remains 
one of the unaddressed issues in existing methods, which also motivates our present work.


\noindent \textbf{Present work}: In this paper, we propose a Heterogeneous Dynamical Graph Neural Networks (HDGNN)-based approach to study the dynamic evolving process of scientific impact while capturing rich semantics embedded in bibliographic graphs. HDGNN bridges the gap between dynamical GNNs~\cite{Trivedi2019,Manessi2020} and heterogeneous information network (HIN) embedding~\cite{zhang2019heterogeneous,Lu2019,Shi2018}, which have 
largely been studied independently in the prior works. 
HDGNN learns academic graph representation with a heterogeneous GNN that aggregates neighboring features of nodes with a newly designed weighted contextualized node selection strategy and temporal-attentive representation network, while preserving the unevenly distributed scientific impact of nodes. 
It also captures the dynamic evolution of nodes and the temporal dependencies among authors/papers, by encoding temporal cascading information into node representations which, in turn, sheds light on the underlying mechanism that accumulates the impact for both authors and papers.

Our main contribution 
is two-fold: (1) 
We study scientific impact quantification problem from the view of heterogeneous graph learning compared to prevalent homogeneous graph/cascade learning models~\cite{li2017deepcas,chen2019cascn}, which allows us to capture complex and rich interactions among different types of nodes. (2) 
We extend heterogeneous graph learning with a temporal horizon, enabling us to address the dynamic prediction problem -- compared to existing HIN embedding works, which mainly focus 
on graph representation learning and/or several relevant \textit{static} tasks such as link prediction and node classification~\cite{zhang2019heterogeneous,Lu2019}.



\section{Preliminaries}
\label{sec:preliminaries}

We now introduce the necessary background and formally define our problem settings. 

Consider \textit{papers} and \textit{authors} as two independent sets of entities, denoted as $P$ and $A$, respectively. For each paper $p$, $p.t$ indicates the time since its first publication. For each author $a$, $a.t$ represents how many years this author has been publishing papers. Let $t_r$ be the reference time and $t_p$ be the prediction time, $c_{p}^{t}$ be 
the number of citations of paper $p$ at time $p.t$, and $c_{a}^{t}$ 
the number of citations of author $a$ at time $a.t$. 
The scientific impact predictions 
for papers and authors can be defined as regression problems as follows: 

\begin{problem}[Scientific impact prediction for papers]
Given $N$ papers $\{p_i\}_{i\in N}$, for each paper $p_i$ and its associated observations at time $p.t_r$, we aim to predict its total number of citations $c_{p_i}^{t_p}$ at prediction time $p.t_p$, i.e., how many times this paper has been cited since publication. 
\end{problem}

\begin{problem}[Scientific impact prediction for authors]
Given $M$ authors $\{a_i\}_{i\in M}$, for each author $a_i$ and its associated observations at time $a.t_r$, we aim to predict its total number of citations $c_{a_i}^{t_p}$ at prediction time $a.t_p$, i.e., how many times this author has been cited since her first publication. 
\end{problem}

Given the above 
we build an academic heterogeneous graph $\mathcal{G} = (\mathcal{V}, \mathcal{E}, \mathcal{A}, \mathcal{R}, \mathbf{C}_\mathcal{V})$, where $\mathcal{V}$ denotes 
the set of nodes and $\mathcal{E}$ is the set of weighted and directed edges 
indicating node relations. For each node in $\mathcal{V}$, it is associated with a node type in $\mathcal{A}$ (
we consider $|\mathcal{A}| = 3$ types of nodes: paper, author, and venue). Edges in $\mathcal{E}$ are described by $7$ different types defined in $\mathcal{R}$: author \textit{writes} paper, author \textit{collaborates with} author, author \textit{publishes in} venue, author \textit{cites} paper, paper \textit{is published in} venue, paper \textit{cites} paper, and paper \textit{cites} author(s). Additionally, node features are represented by $\mathbf{C}_\mathcal{V}$, including content of papers, profiles of authors, etc. 

Suppose we have $N$ papers and $M$ authors. 
During an observation window each paper or author can be cited by other papers/authors. Then, the sequence of citations can be represented as $[(p_j, t_j)]_j (t_j \leq t_r$), i.e., 
cascades. 
Given the exact number of citations $c^{t_p}$ at prediction time $t_p$ for a particular paper/author, the scientific impact prediction problem can be solved by optimizing the mean square error loss 
between predicted number of citations $\hat{c}^{t_p}$ and the true number $c^{t_p}$.

\section{Methodology: HDGNN}
\label{sec:model}
We now present our proposed model \M, which 
consists of two main building blocks: (i) heterogeneous representation learning via Graph Neural Networks (GNNs); and (ii) temporal paper sequence modeling and author aggregating via Recurrent Neural Networks (RNNs). For simplicity, here we use \textit{scientific impact prediction for papers} as an illustrative scenario, 
with a note that the results can be easily generalized to the scenario of \textit{scientific impact prediction for authors} (we show the prediction results for both settings in Section~\ref{sec:experiments}).

\subsection{Heterogeneous Graph Representation}
\label{subsec:deep-hetero}

The first part of \M~is to learn representation of nodes in $\mathcal{G}$. Specifically, for a paper node $p$, author node $a$, and venue node $v$ -- given heterogeneous neighbors in a non-Euclidean graph structure -- 
we learn a low-dimensional node embedding $E(p/a/v)$ via a mapping function $f: p/a/v$ $ \to E(p/a/v) \in \mathbb{R}^{d_E}$ 
and the embedding preserves neighboring proximity. Towards that, we borrow the idea of random walk with restart \cite{tong2006fast} and a deep neural network architecture \cite{lecun2015deep} from \cite{zhang2019heterogeneous} to model the heterogeneous graph representation learning.

\begin{figure}[htbp]
    \centering
    \includegraphics[width=0.49\textwidth]{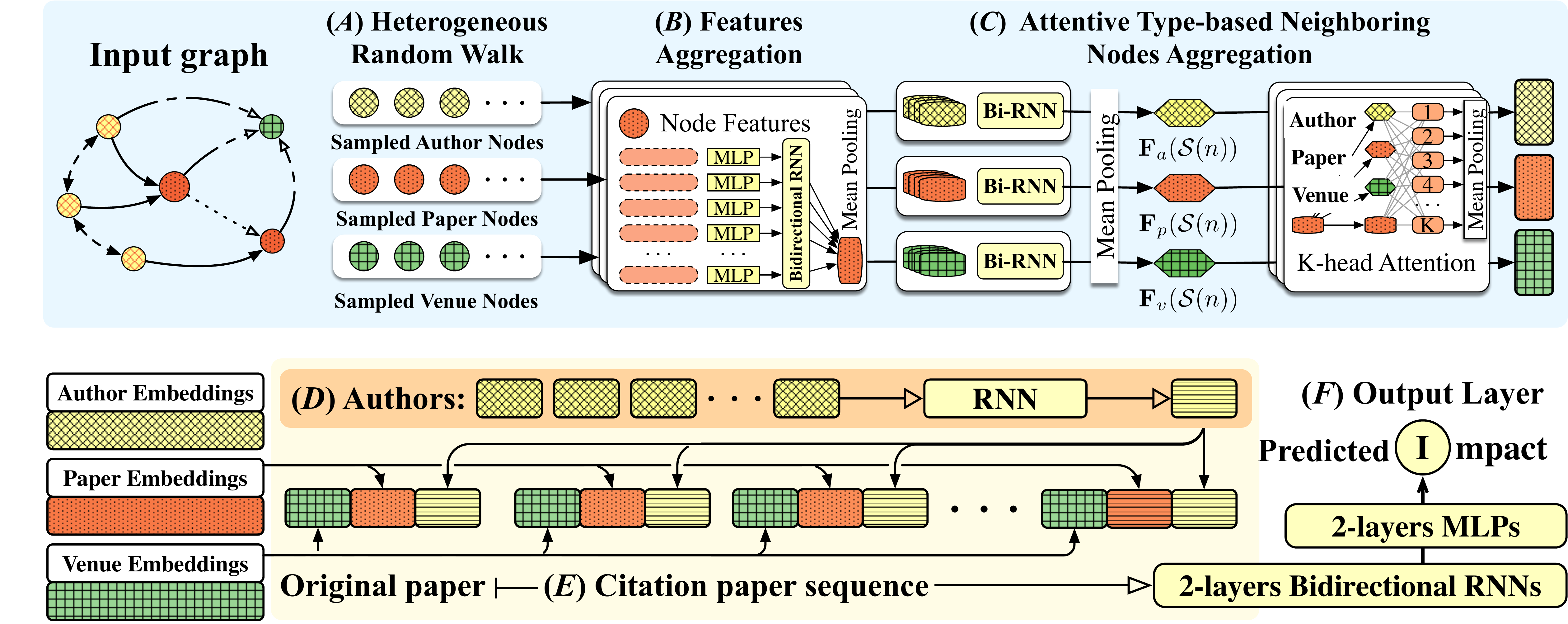}
        \caption{The overall architecture of 
        \M: 
        \textbf{\textit{(A)}}: Walk heterogeneous nodes by using a weighted contextualized node selection strategy based on random walk with restart; \textbf{\textit{(B)}} and \textbf{\textit{(C)}}: Aggregating node features and heterogeneous neighbors of nodes with bidirectional RNNs and multi-head attention mechanism; 
        \textbf{\textit{(D)}}: Multiple author aggregation; \textbf{\textit{(E)}}: Sequential citation aggregation; \textbf{\textit{(F)}}: Scientific impact predictor with RNNs and MLPs.
        }
    \label{fig:model}
    \vspace{-0.3cm}
\end{figure}

\paragraph{Heterogeneous neighboring node sampling.} Given a node $n$ (paper/author/venue) in graph $\mathcal{G}$, the distribution of its neighboring nodes may be 
highly skewed, i.e., some nodes connect to a large number of other nodes (e.g., those highly cited papers/authors) while most of them only have a few 
neighbors, greatly following the heavy-tailed distribution of citations~\cite{Fortunato2018}. High impact journals, productive authors, or influential papers, often have higher degrees compared to other majorities. To accommodate this factor into our model, we design a weighted contextualized node selection strategy based on random walk, which is more suitable for capturing scientific impact and imbalanced distribution of nodes in the heterogeneous academic graph. Specifically, for each current step, a given node $n$
either 
returns to the previous node with probability $q$, or jumps to the next neighbor node with probability $1-q$. Let $\mathcal{N}(n)$ be the set of $n$'s neighbors. 
Then the node $n$ has a probability $1-q$ to select one of its neighbors $\mathcal{N}(n)$ based on node types $\mathcal{A}$, edge types $\mathcal{R}$, and node/edge characteristics. Specifically, the probability to walk to the next node $m$ from $n$ is:
\begin{equation}\label{eq:rwr}
\Pr(m|\mathcal{N}(n), \mathcal{G}) = \begin{cases}
     (1-q) \alpha D^\alpha(m), & \text{if } \mathcal{A}_m \text{ is paper} \\
     (1-q) \beta D^\beta(m), & \text{if } \mathcal{A}_m \text{ is author} \\
     (1-q) \gamma D^\gamma(m), & \text{if } \mathcal{A}_m \text{ is venue}
\end{cases}
\end{equation}
where $D^\alpha(*), D^\beta(*), D^\gamma(*)$ are influence functions measuring node influence from various factors, e.g., arrival time, node degrees, pagerank scores, and similarities, according to the node type $\mathcal{A}_m$ or edge type $\mathcal{R}_e$.

Through running random walks iteratively, we can sample a fixed number of nodes for each node type in $\mathcal{A}$, resulting in three sets denoted as: 
$\mathcal{S}(p)$, $\mathcal{S}(a)$, and $\mathcal{S}(v)$, respectively. Note that we consider edge directions, 
weights, and node degrees when sampling representative heterogeneous neighbors.


\paragraph{Aggregating node features.} After sampling the neighbors for each node, we utilize Bidirectional Gated Recurrent Units (Bi-GRUs)~\cite{chung2014empirical} to model the dependencies among the nodes' content features. 
Assuming that there are $k$ content features for one specific type of nodes, the feature aggregation can be formalized as:
\begin{align}\label{eq:bi-gru}
    \mathbf{F}(n) &= \frac{1}{k} \sum_{i=1}^{k} \left(\overrightarrow{\text{GRU}}(\mathbf{h}_n^i) \left|\right| \overleftarrow{\text{GRU}}(\mathbf{h}_n^i)\right), \\
    \mathbf{h}_n^i &= \text{MLP}(\mathbf{C}_n^i), \text{ for } i = 1, 2, \dots, k
\end{align}
where $\mathbf{F}(n) \in \mathbb{R}^{d_n}$ is the aggregated embedding of node $n$ computed by mean pooling; $||$ denotes the concatenation operation; $\mathbf{C}_n$ are $k$ heterogeneous node content features and $\mathbf{h}_n^i \in \mathbb{R}^{d_\mathbf{h}}$ is the output of the Multi-Layer Perceptron (MLP). In practical applications, various content features can be used here to enhance the model learning ability -- e.g., meta-data and the text of papers (title, abstract, main body), illustrations/figures 
historical publications of authors/venues, profiles and honors of authors.

\paragraph{Aggregating heterogeneous neighbors.} After aggregating node content features, for each node $n$ in the graph $\mathcal{G}$ we have its corresponding aggregated features $\mathbf{F}(n)$. Then we are ready to use a type-based RNN to aggregate embeddings of the neighbors in $\mathcal{S}(n)$. For each node type in $\mathcal{A}$ (in our case the paper/author/venue), $\mathcal{S}_{p/a/v}(n)$ is the homogeneous type-specific neighboring set of node $n$ and $\text{RNN}_{p/a/v}$ is a type-specific aggregator. More specifically, \M~utilizes another Bi-GRU for modeling $n$'s neighbors:
\begin{align}\nonumber
    \mathbf{F}_{\mathcal{A}_n}(\mathcal{S}_{\mathcal{A}_n}(n)) &= \frac{\sum_{i=1}
    \left( \overrightarrow{\text{GRU}}(\mathbf{F}(i)) ||
    \overleftarrow{\text{GRU}}(\mathbf{F}(i)) \right)}{|\mathcal{S}_{\mathcal{A}_n}(n)|}, \\\label{eq:neighbor-aggre}
    \text{ for } i &= 1, 2, \dots, {|\mathcal{S}_{\mathcal{A}_n}(n)|} 
\end{align}
where $\mathbf{F}_{\mathcal{A}_n}(\mathcal{S}(n)) \in \mathbb{R}^{d_s}$ is the output embedding from the homogeneous neighboring set $\mathcal{S}_{\mathcal{A}_n}(n)$, and $d_s$ is the dimension of aggregated neighboring embeddings of node $n$. 

In HDGNN we use deterministic neural networks, bidirectional RNN, and mean pooling as aggregators of node's content along with node's neighbors. Alternatively, other types of aggregators, e.g., last hidden state of RNNs, CNNs, max pooling, can be used 
(cf.~\cite{zhang2019heterogeneous,hamilton2017inductive}). 

\paragraph{Multi-head attention for type-based neighbors.} With each of the type-based neighboring aggregators in hand, we are able to combine them using multi-head attention mechanism \cite{velivckovic2018graph}:
\begin{gather}\nonumber
    \alpha_{i} = \frac{
        \exp \left( \text{LeakyReLU}(u^T[\mathbf{F}(n) || \mathbf{F}_i^{\mathcal{S}})] \right)
    }{
        \sum_{j \in \mathcal{S}'(n)} \exp \left( \text{LeakyReLU} (u^T[\mathbf{F}(n) || \mathbf{F}_j^{\mathcal{S}}]) \right)
    }, \nonumber \\\nonumber
    \mathcal{S}'(n) = \mathbf{F}(n) \cup \{\mathbf{F}_j^\mathcal{S}\}_{j\in \mathcal{S}(n)}, \\\label{eq:attention}
    E(n) = \frac{1}{K} \sum_{i=K} \sum_{\mathbf{F}_i(n) \in \mathcal{S}'(n)} \alpha_i \mathbf{F}_i(n)
\end{gather}
where $E(n)\in \mathbb{R}^{d_E}$ is the learned embedding of node $n$, LeakyReLU is the activation function, $||$ denotes the concatenation operation, $u$ is the attention parameter, and $K$ is the number of attention heads. Here, $\mathbf{F}(n)$ and $\mathbf{F}_j^\mathcal{S} = \mathbf{F}_{\mathcal{A}_j}(\mathcal{S}_{\mathcal{A}_j})$ are computed by Eq.~\eqref{eq:bi-gru} and Eq.~\eqref{eq:neighbor-aggre}, respectively. 

\subsection{Citation Cascading and Author Aggregation}
\label{subsec:temporal}

The second part of \M~is to model the cascading behavior of the papers/authors. Here we consider each paper $p$ as an independent entity. Recall that $p.t_0$ is the publication time of $p$, $p.t_r$ is the reference time, $\{(p_j, t_j)\}_{j}$ is the set of citation papers of $p$ published at time $t_j$ during the observation window $[p.t_0, p.t_r] (t_j\leq p.t_r)$. Since we already obtained embeddings of the papers $E(p)$, authors $E(a)$, and venues $E(v)$ (cf.~Eq.~\eqref{eq:attention}), we now separately model authors of a paper and the paper itself by feeding them into RNNs. 

\paragraph{Multi-author aggregation layer.} Note that 
each citing paper $p_j$ of a 
given/original paper $p$, 
may contain multiple authors (in our dataset the mean number of authors per paper is 3.438 and the max is 25). We sequentially pipeline the author embeddings 
into a GRU and then use the last hidden state $\mathbf{h}_{p_j}^a$ as the representation of $p_j$'s authors. 

\paragraph{Sequential citation aggregation layer.} After author aggregation, for each paper $p_j$, 
we have its own embedding $E(p_j)$, the corresponding venue embedding $E(v_j)$,
and the aggregated author embeddings $E(\mathbf{a}_j) = \mathbf{h}_{p_j}^a$. 
We then use a two-layer Bi-GRU to sequentially aggregate the citing papers ordered by their publishing time $t_j$, where each citing paper $p_j$ is modeled as the combination of paper, authors, and venue. The rational{\`e} 
is that we expect to capture temporal dependencies among citing papers, which, as we will show in the experiments, is superior to 
other aggregators such as sum or max pooling~\cite{hamilton2017inductive}. The overall architecture of the citation aggregation can be formalized as:
\begin{align}\label{eq:citation-aggre}\nonumber
    \mathbf{E}(p_j) &= (E(p_j) || E(\mathbf{a}_j) || E(v_j)), \\\nonumber
    \mathbf{h}_j^1 &= (\overrightarrow{\text{GRU}}(\mathbf{E}(p_j)) || \overleftarrow{\text{GRU}}(\mathbf{E}(p_j))), \\
    \mathbf{h}_j^2 &= (\overrightarrow{\text{GRU}}(\mathbf{h}_i^1) || \overleftarrow{\text{GRU}}(\mathbf{h}_i^1))
\end{align}
where $\mathbf{h}_j^2 \in \mathbb{R}^{d_{\mathbf{h}^2}}$ is the $j$-th hidden state of the second layer of Bi-GRU.
Here we concatenate the last hidden state of Bi-GRU
as the final output representation of paper, and then make use of it to predict the scientific impact of $p$.

\begin{figure*}[t]
    \centering
    \includegraphics[width=0.9\textwidth]{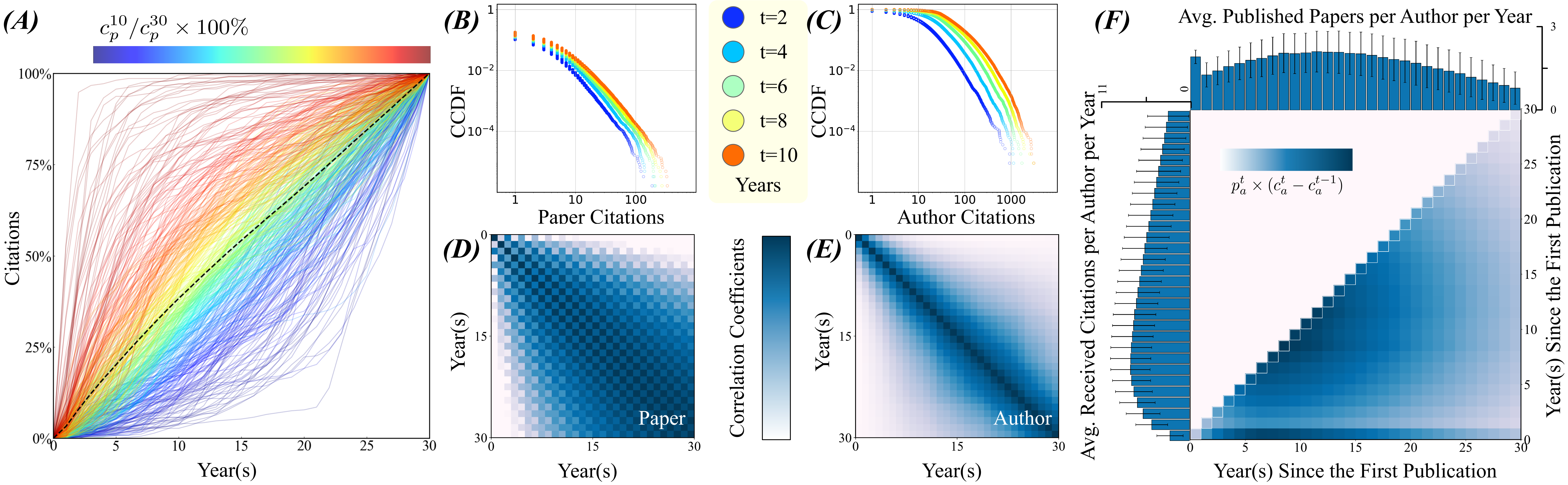}
    \caption{Dataset statistics: \textbf{\textit{(A)}}: 509 papers 
    with more than 200 citations after 30 years since publication before 1987. Lines represent normalized citation growth trends, 
    line colors 
    indicate citation rank of papers at the tenth year, i.e., $c_{a}^{10}/c_{a}^{30}\times 100\%$; 
    dashed line denote the mean values. \textbf{\textit{(B)}} and \textbf{\textit{(C)}}: Complementary cumulative distribution function (CCDF) of paper citations and author citations, respectively. \textbf{\textit{(D)}} and \textbf{\textit{(E)}}: Pearson correlation coefficients of paper citations and author citations over 30 years, respectively. The $(i,j)$ block in heatmap represents the correlation between $i$-th year's and $j$-th year's cumulative citations for all papers/authors. \textbf{\textit{(F)}}: The value of the $i$-th diagonal block of the heatmap is 
    $p_{a}^t\times (c_{a}^t-c_{a}^{t-1})$, i.e., the average number of papers each author published at the $i$-th year multiplied by the average number of citations each author received at the $i$-th year; 
    Top histogram: 
    average published papers per author per year, 
    Left histogram: 
    average citations authors received per year 
    (errorbars are standard deviations).}
    \label{fig:data} 
\end{figure*}
%

\paragraph{Output and model training.} The output of \M~is the predicted citation number $c_p^{t_p}$ of a paper $p$. We use a two-layer of MLPs with GeLU activation \cite{hendrycks2016bridging}.
The training losses of 
graph representation and impact prediction are respectively defined as:
\begin{align}\label{eq:loss1}
    \mathcal{L}_1(\Theta_1) &= \arg \max_{\Theta_1} \prod_{n\in \mathcal{V}} \prod_{\mathcal{A}_n}\prod_{n_c\in \mathcal{N}_{c}} \Pr (n_c|n;\Theta_1),  \\\label{eq:loss2}
    \mathcal{L}_2(\Theta_2) &= \frac{1}{N_T}\sum_{i=1}^{N_T} (\hat{c}_{p_i}^{t_p} - c_{p_i}^{t_p})^2 
\end{align}
where $\mathcal{N}_c(n)$ is the type-based neighboring set of node $n$, $\Pr (n_c|n;\Theta_1)$ is the conditional probability, 
$N_T$ is the number of training samples, and $\hat{c}_{p_i}^{t_p}$ is the predicted number of citations of paper $p_i$ at time $t_p$. 

As for scientific impact prediction for scholars, the general training process is similar, except that Eq.~\eqref{eq:loss2} is alternatively defined as: $
    \mathcal{L}_2(\Theta_2) = \frac{1}{M_T} \sum_{i=1}^{M_T} (\hat{c}_{a_i}^{t_p} - c_{a_i}^{t_p})^2, 
$where $\hat{c}_{a_i}^{t_p}$ is the predicted number of citations for author $a_i$.


\section{Experiments}
\label{sec:experiments}
We evaluate HDGNN and several baselines on two scientific impact predictions -- 
paper and author citations, respectively. 

\paragraph{Dataset.} The evaluations were 
performed on American Physical Society (APS) dataset\footnote{\url{https://journals.aps.org/datasets}}. The APS dataset contains over 422K academic papers on 17 venues and 54M citations among papers between 1893 and 2017. 
The constructed heterogeneous graph of APS contains 616,316 papers, 430,950 authors, and 17 venues. 
For edges we have: author \textit{writes} paper (2.9M), author \textit{collaborates with} author (2.8M), author \textit{publishes in} venue (0.6M), author \textit{cites} paper (20.5M), paper \textit{cites} paper (7.3M), paper \textit{cites} author (4.8M), paper \textit{is published in} venue (0.6M). For papers (authors) in the dataset, we select 20 years as the prediction time $p.t_p$ ($a.t_p$). Thus, we only consider the papers published before 1997, to 
ensure that each paper has at least 20 years to grow its citations count. In the same way, selected authors are required to start their research career no later than 1997. We set reference time to 2 years and we 
note that papers/authors whose citations are less than 10 during the observation window are filtered out. The settings of prediction for authors are as the same as that for papers. After the preprocessing, we have a 
total of 11,475 papers and 14,318 authors. We use 50\% of them for training, 25\% for validation and the rest 25\% for testing. 
Figure~\ref{fig:data} shows the statistics of APS dataset.

\paragraph{Baselines.} The baselines used for 
comparison 
include feature-oriented models, graph embedding models, graph neural networks, as well as the state-of-the-art information cascade popularity prediction models.

\noindent $\bullet$ \textbf{Uniform} -- for all papers/authors, we always predict their impact as a fixed number, uniformly searched from the minimum $\log c_{p/a}^{t_p}$ to maximum $\log c_{p/a}^{t_p}$ with a step of $0.001$. 

\noindent $\bullet$ \textbf{Feature} -- we extract features into a linear regression model: observed citations $c_{p/a}^{t_r}$, mean arrival time,
and degrees of nodes. 
We use observed citation $c_{p/a}^{t_r}$ (i.e., Feature-$c^{t_r}$) as a simple baseline. 


\noindent $\bullet$ \textbf{DeepCas} \cite{li2017deepcas} -- is a deep learning based prediction model utilizing DeepWalk for graph embedding and RNNs for cascade modeling and predicting. 

\noindent $\bullet$ \textbf{DeepHawkes} \cite{cao2017deephawkes} -- makes use of Hawkes point processes and neural networks for cascade prediction. 

\noindent $\bullet$ \textbf{CasCN} \cite{chen2019cascn} -- utilizes GCN~\cite{Kipf2017} and LSTM to model the structural and temporal information of cascades. 

\paragraph{Variants of HDGNN.} In order to compare other graph representation frameworks with our proposed \M, we select following 10 models to replace the first part of \M~as variants, including homogeneous or heterogeneous methods, skip-gram based or matrix factorization based methods: \textbf{DeepWalk} \cite{perozzi2014deepwalk}, \textbf{LINE} \cite{tang2015line}, \textbf{Metapath2Vec} \cite{dong2017metapath2vec}, \textbf{ProNE} \cite{zhang2019prone}, together with graph neural network \textbf{GraphSAGE} \cite{hamilton2017inductive} and \textbf{HetGNN} \cite{zhang2019heterogeneous}. Besides, we substitute the RNN aggregator with max pooling or sum pooling as two additional variants, denoted as \textbf{\M-MaxP} and \textbf{\M-SumP}. To evaluate the impact of author/venue embeddings, we separately remove the author part or venue part in Eq.~\eqref{eq:citation-aggre} as \textbf{\M-NoAuthor} and \textbf{\M-NoVenue}. 






\paragraph{Metrics.} We use two widely used evaluation metrics \cite{zhao2015seismic,shen2014modeling,cao2017deephawkes}, i.e., mean square logarithmic error (MSLE) and accuracy (ACC):

\noindent $\bullet$ MSLE: $\frac{1}{N_t}\sum_{i=1}^{N_t} (\log \hat{c}_{p_i/a_i}^{t_p} - \log c_{p_i/a_i}^{t_p})^2$;

\noindent $\bullet$ ACC: $\frac{1}{N_t}\sum_{i=1}^{N_t} \mathbf{1}(0.5 * c_{p_i/a_i}^{t_p} \leq \hat{c}_{p_i/a_i}^{t_p} \leq 1.5 * c_{p_i/a_i}^{t_p})$;

\noindent where $\mathbf{1}(\cdot)$ is the indicator function, $N_t$ is the test sample size.


\begin{table}[t]
    \centering
    \small
    \begin{tabular}{lcccc}
    \toprule
        \multirow{2}{*}[-3pt]{\textbf{Model}}              & \multicolumn{2}{c}{\textbf{Papers}}       & \multicolumn{2}{c}{\textbf{Authors}} \\ \cmidrule(lr){2-3} \cmidrule(lr){4-5}
                                    & \textbf{MSLE}   & \textbf{ACC}      & \textbf{MSLE}   & \textbf{ACC}      \\ \midrule
        Uniform                     & 0.588 & 49.70\%   & 1.102 & 36.37\%   \\ 
        Feature-$c^{t_r}$           & 0.401 & 58.35\%   & 0.939 & 39.16\%   \\
        Feature                     & 0.361 & 58.77\%   & 0.832 & 40.19\%   \\
        DeepCas                     & 0.349 & 58.22\%   & 0.787 & 41.45\%   \\
        DeepHawkes                  & 0.328 & 59.91\%   & 0.725 & 42.38\%   \\
        CasCN                       & 0.310 & 61.71\%   & 0.692 & 44.03\%   \\ \midrule
        DeepWalk                    & 0.288 & 68.89\%   & 0.627 & 49.09\%   \\
        LINE                        & 0.281 & 68.70\%   & 0.614 & 47.88\%   \\
        Metapath2Vec                & 0.294 & 66.38\%   & 0.642 & 49.72\%   \\
        GraphSAGE                   & 0.309 & 64.97\%   & 0.675 & 48.46\%   \\ 
        HetGNN                      & 0.292 & 65.79\%   & 0.607 & 51.06\%   \\ 
        ProNE                       & 0.297 & 66.60\%   & 0.635 & 47.65\%   \\
        \midrule
        \M-MaxP                     & 0.358 & 64.22\%   & 0.810 & 42.71\%   \\ 
        \M-SumP                     & 0.280 & 69.90\%   & 0.749 & 44.10\%   \\
        \M-NoAuthor                 & 0.279 & 69.05\%   & 0.605 & 50.00\%   \\
        \M-NoVenue                  & 0.290 & 67.10\%   & 0.651 & 48.26\%   \\
        \midrule
        \textbf{\M} & \textbf{0.268} & \textbf{69.77\%} & \textbf{0.590} & \textbf{51.62\%} \\
    \bottomrule
    \end{tabular}
    \caption{Performance comparison: prediction for papers/authors. Bold: \textit{t}-test compared to the best baseline ($p<0.005$).} 
    \label{tab:performance}
\end{table}

\paragraph{Experimental settings.} For all graph representation baselines, we set the embedding dimension to 128. Random walk restart probability $q$ is 0.5, walk length is 30, and number of walks for each node equals to 5.
For type specific parameters $D^\alpha(*), D^\beta(*), D^\gamma(*)$, we use node in-degree and edge weights as a proxy of node influence. For \M~and its variants, the learning rate is chosen from $\{1, 10^{-1}, \dots, 10^{-5}\}$, and the node embedding size is $128$. The length of citation sequence of all methods (whether RNN, LSTM or GRU) is set to $100$ -- i.e., the max number of citation sequence. For papers/authors whose length is more than $100$, we only select their first $100$ citations (as for author sequence, the length of RNN is set to $6$). The units 
are set to 128 and 64 in two-layer Bi-RNNs, and to 64 and 32 
in two-layer MLPs. 
For feature aggregation RNNs, we use paper title embeddings pre-trained via BERT \cite{Devlin2019}, and node embeddings pre-trained via DeepWalk. All the 
other hyper-parameters of baselines are set to their default values. Performance results are reported with early stopping on validation loss of 10 epoch patience. 
The source code of \M~is released at \url{https://github.com/Xovee/hdgnn}. 


\paragraph{Prediction performance.} We show the performance of all the models in Table~\ref{tab:performance}, 
and we observe that: 

\noindent (1) 
HDGNN outperforms all the other methods in both paper and author impact prediction. This result demonstrates 
the effectiveness of learning interactions among heterogeneous nodes with the proposed heterogeneous information aggregation, which can be further verified by the fact that both feature-based models and homogeneous cascade prediction methods do not show comparable performance. Previous popularity prediction methods, e.g., DeepCas, DeepHawkes and CasCN, do not distinguish the type of nodes and therefore fail to model their complex and meaningful interactions.

\noindent (2) 
Author impact prediction is much harder than that of papers. As shown in \textbf{\textit{(B)-(E)}} in Figure~\ref{fig:data}, the citation number of authors is higher than that of papers by orders of magnitude, as well as the coefficients of correlation between observed and future citations. In fact, in settings of two year observation, the proportion of average observed citations $c_p^2$ to $c_p^{20}$ is about 9.1\% for authors. In contrast, the proportion for papers is 34.6\% (cf.~\textbf{\textit{(A)}} in Figure~\ref{fig:data}), which explains why prediction for authors' impact is more difficult -- i.e., largely due to insufficient observations and enormous variability in scholars' productivity~\cite{Clauset2017} (cf.~\textbf{\textit{(F)}} in Figure~\ref{fig:data}). In addition, paper citation is strongly correlated to the factors such as the citations a paper has gained and the importance of publication venue (e.g., journal impact factor), which can be easily modeled in the graph with node attributes. In contrast, scholars' impact is far more unstable due to implicit factors such as funding scheme, tenure, gender issues 
-- all of which need to be quantified with external high-resolution data repositories.



\paragraph{Ablation study.} We now investigate the effect of important modules in HDGNN. Firstly, the information aggregation mechanism used in HDGNN is better than other graph embedding techniques including two heterogeneous network embedding methods, i.e., Metapath2Vec and HetGNN -- because of the more complex relations considered in our model and the benefit of considering temporal dependencies between citation sequences and/or author sequences. For example, \M~models 7 types of relations among nodes, whereas HetGNN, in contrast, only considers 3 edge types. 

Additionally, 
the publication venue plays a vital role in predicting the impact of an author or a paper. This is demonstrated 
by the significant performance degradation 
after removing venue embedding in Eq.~\eqref{eq:citation-aggre}. Authorship, surprisingly, is less important than the journal that a paper published in, though masking the authorship information may slightly degrade the prediction performance. As for aggregation choices, both max pooling and sum pooling are inferior to the RNN aggregator used in HDGNN, due to their lack of sequential dependencies, which are important for evolving trend prediction.



\begin{figure}[t]
    \centering
    \includegraphics[width=.48\textwidth]{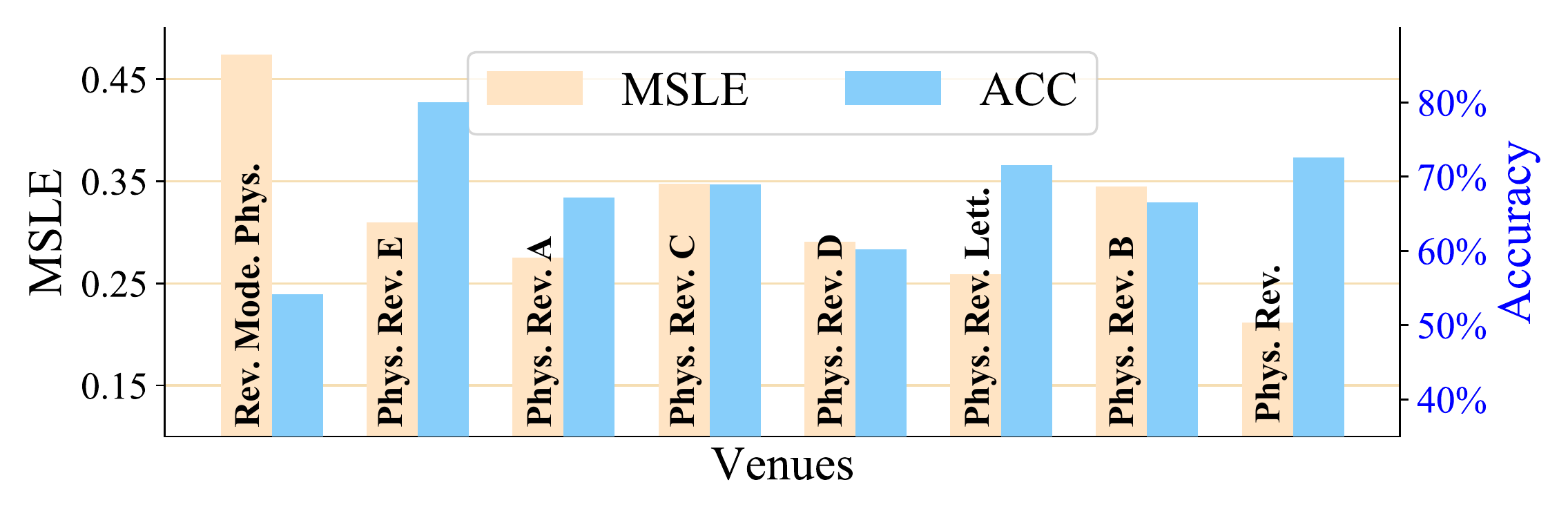}
    \caption{Performance of HDGNN on 8 representative venues.}
    \label{fig:venues} 
    \subfloat[Papers]{
		\includegraphics[width=0.23\textwidth]{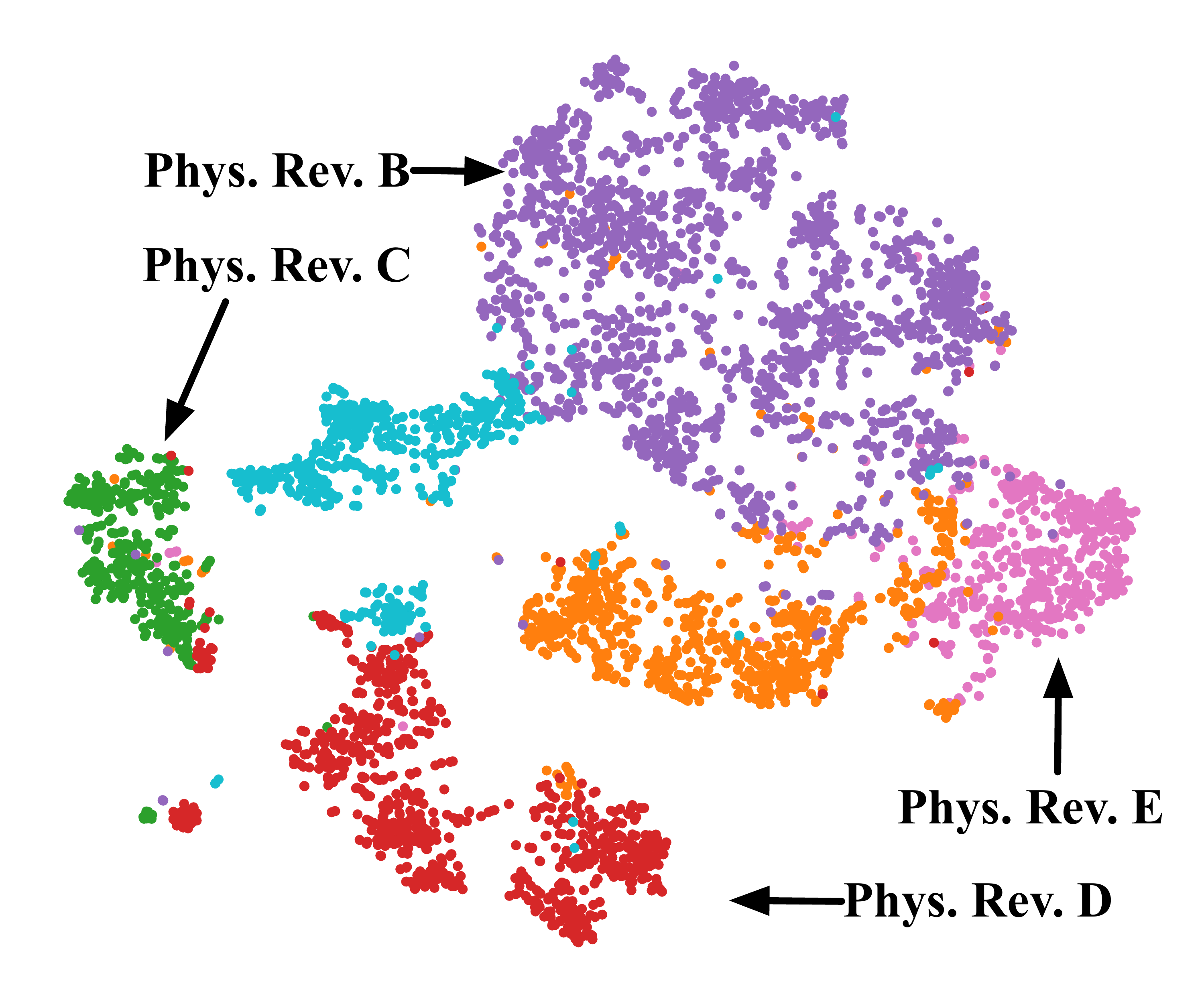}
	}
	\subfloat[Papers]{
		\includegraphics[width=0.23\textwidth]{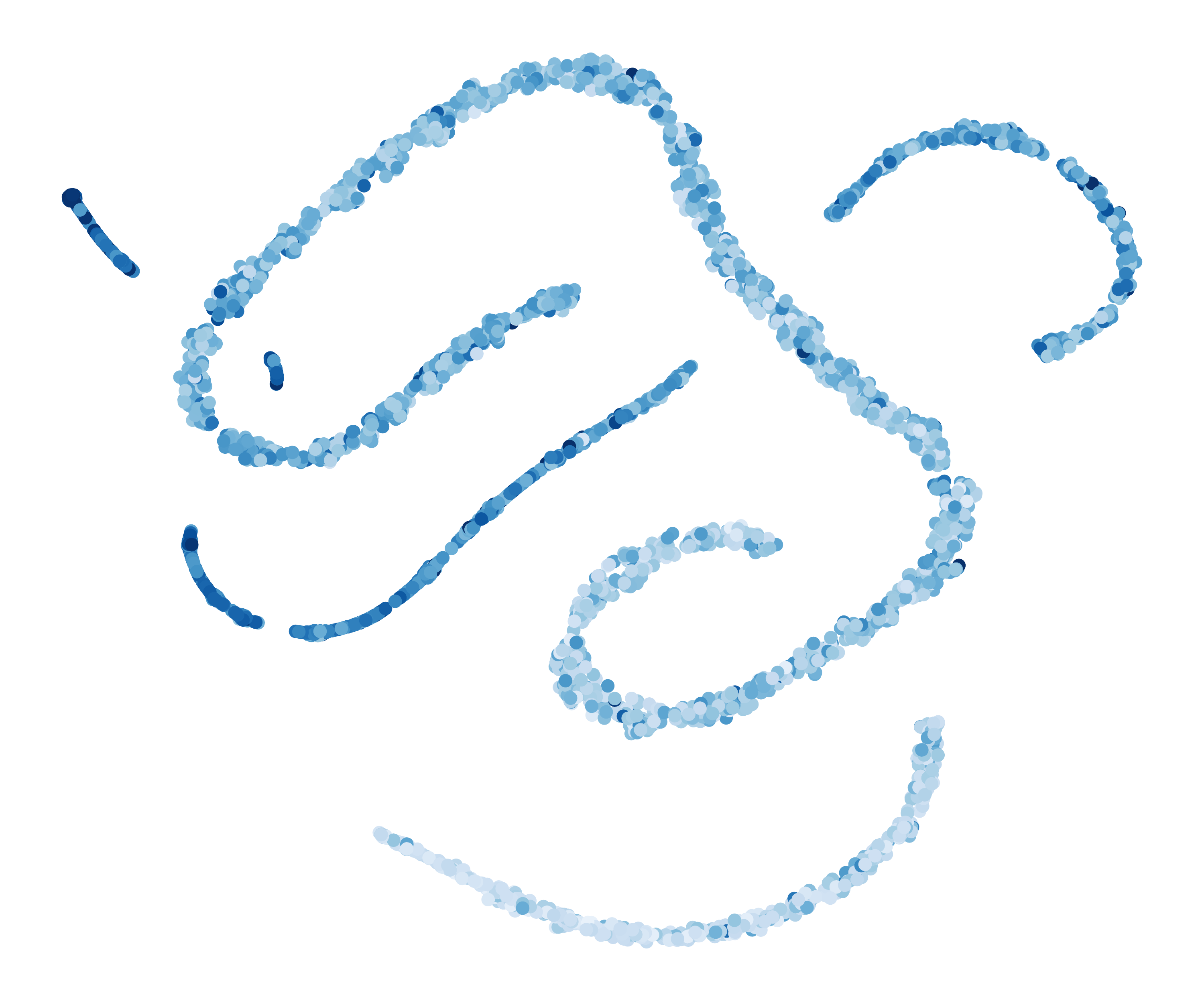}
	}
	
	\subfloat[Authors]{
		\includegraphics[width=0.23\textwidth]{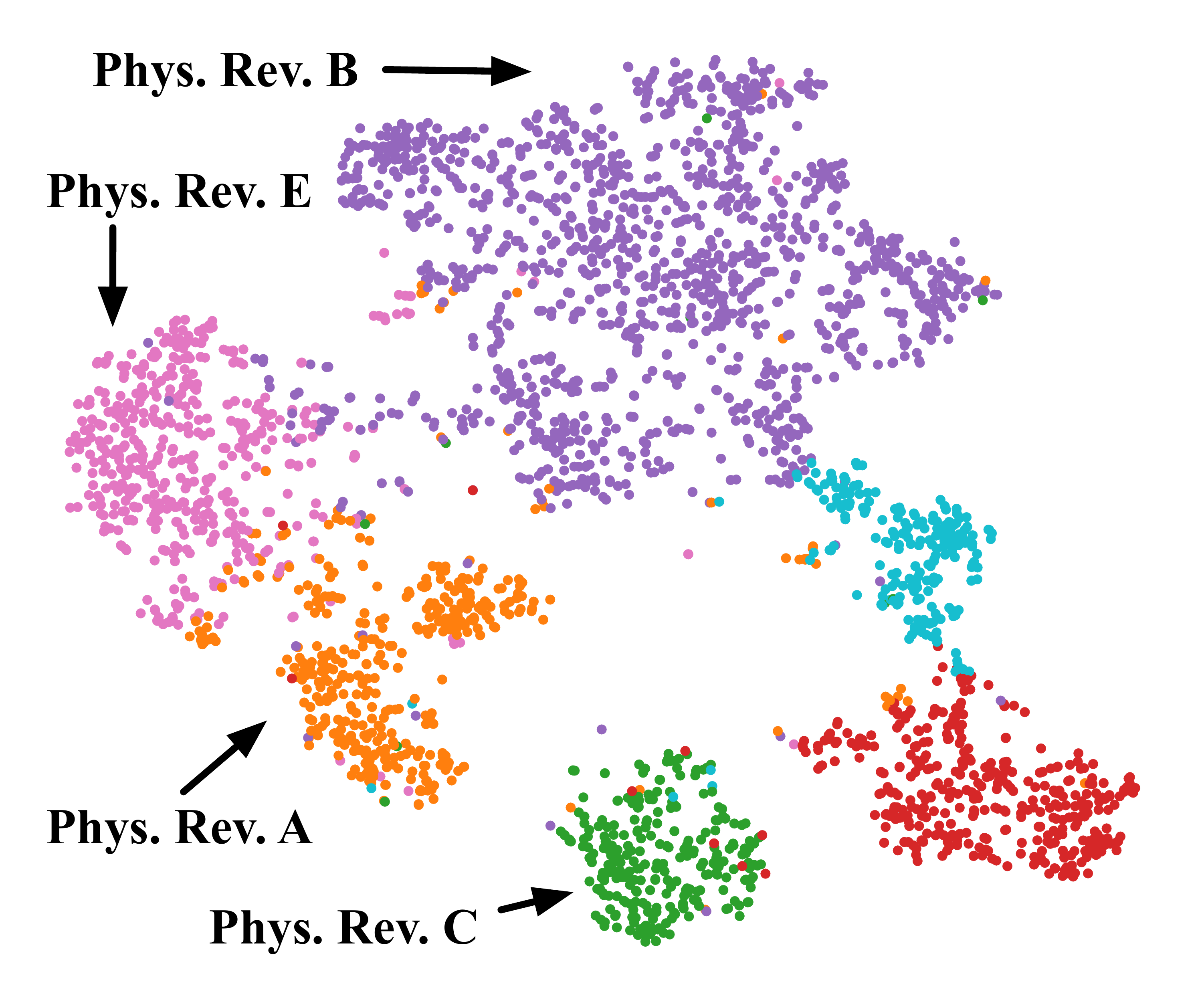}
	}
	\subfloat[Authors]{
		\includegraphics[width=0.23\textwidth]{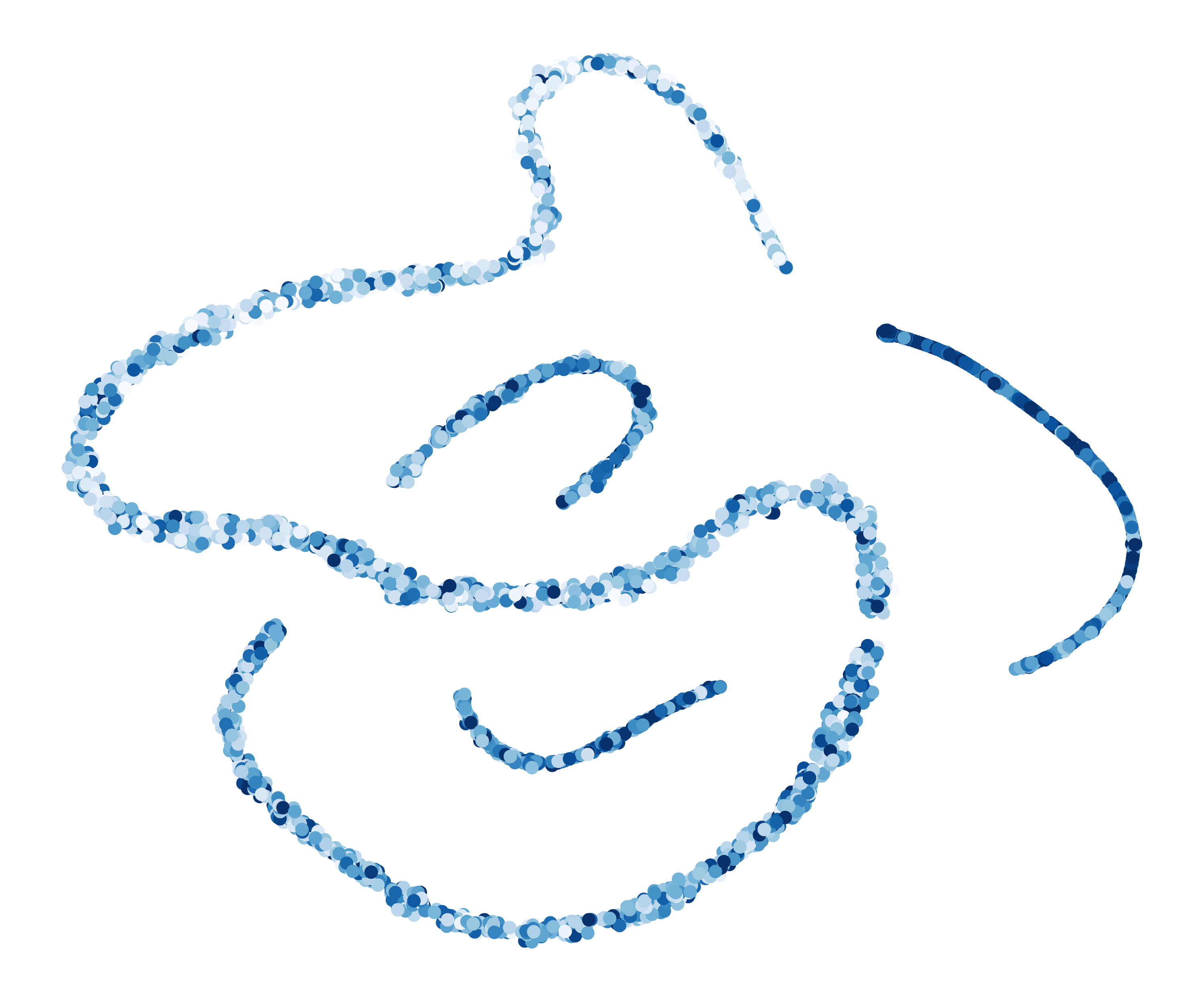}
	}
    \caption{Plot latent space of 6 venues after mapping to 2D with t-SNE (best viewed in color). (\textbf{\textit{A}}) and (\textbf{\textit{C}}) show paper/author embeddings retrieved from the heterogeneous graph representation (i.e., $E(n)$, cf.~Section~\ref{subsec:deep-hetero}) -- colors are specified by venues; (\textbf{\textit{B}}) and (\textbf{\textit{D}}) plot paper/author citation embeddings from the prediction layer (i.e., $\mathbf{h}^2$, cf.~Section~\ref{subsec:temporal}) -- colors are specified by magnitude of citations.}
    \label{fig:t-SNE} 
    \vspace{-0.1cm} 
\end{figure}

\paragraph{Qualitative analysis.} Figure~\ref{fig:venues} shows the prediction results on 8 representative journals -- the lower the MSLE and/or the higher the accuracy, the better performance. The performance of HDGNN varies significantly on different publication venues -- this is natural since venue is a strong indicator for future impact accumulation. In addition, we found that the prediction accuracy is affected by the citation distribution of papers in a journal. 
For example, the standard deviation of 20 year citations of papers (i.e., $c_p^{20}$) on Rev.~Mode.~Phys. is very high (255.02), whereas the value on Phys.~Rev. is significantly less (43.24). This discrepancy also reveals why prediction of papers on Rev.~Mode.~Phys. is more difficult. 


Figure~\ref{fig:t-SNE} plots the latent space learned in HDGNN, where we can observe clear clustering phenomena of author/paper embeddings from \textbf{\textit{(A)}} and \textbf{\textit{(C)}}. It appears that papers published in the same journal tend to cluster together, which also indicates publication venue is an important indicator for scientific impact prediction. In addition, we also visualize a ``crowd effect"~
of high impact papers/authors, as shown in \textbf{\textit{(B)}} and \textbf{\textit{(D)}}. This also implies strong correlations among high impact scholars and papers, e.g., high impact scholars prefer to cite papers from other high impact authors/papers. In other words, there indeed exists a positive feedback loop between high impact papers/scholars. Another interesting result can be visualized is the gradually decaying color of the paper/author citations, implying that heavy-tailed distribution of scientific impact is successfully (to some extent at least) encoded in our model. It could also explain why our dynamic heterogeneous neighboring aggregation with weighted contextualized node selection strategy substantially outperforms other homogeneous and heterogeneous graph embeddings.

\section{Conclusion}
\label{sec:conclusion}

We introduced the 
HDGNN approach for effectively quantifying and predicting the scientific impact of scholars and research 
publications, by bridging the dynamic processes of impact evolution and complex nodes interactions. We presented an efficient network sampling method with the consideration of rich node relations and a temporally attentive neighbor aggregation network to 
model 
the complex and accumulating dynamic processes of scientific impact. Evaluations on a real-world scientific dataset demonstrated the superior performance 
\M~in comparison 
to several state-of-the-art baselines. 
Future work will investigate the impact of cross-institutional collaboration on citations.

\section*{Acknowledgments}

This work was supported by National Natural Science Foundation of China (Grant No.61602097, 61472064, U19B2028 and 61772117), NSF grants III 1213038 and CNS 1646107.

\bibliographystyle{named}
\bibliography{xovee}

\end{document}